# Single-shot image retrieval through a multimode fiber using a genetic algorithm

MICHAEL RUDDLESDEN,[1] JINSHUAI ZHANG,[1] TIANRUI ZHAO,[1] WEN WANG,[1] AND LEI SU[1,*]

[1]*School of Engineering and Materials Science, Queen Mary University of London, London E1 4NS, UK*
*\* l.su@qmul.ac.uk*

**In this letter, we present a genetic algorithm-based approach for image retrieval through a multimode fiber in a reference-less system. Due to mode interference, when an image is illuminated at one side of a multimode fiber, the transmitted light forms a noise-like speckle pattern at the other end. With the use of a prior-measured transmission matrix of the fiber, a speckle pattern is calculated using a random input mask. By optimizing the input mask to achieve a high correlation coefficient of experimental and calculated patterns, the input mask is optimized into an image with high similarity to the original image.**

*OCIS codes: (060.2270) Fiber characterization; (100.3010) Image reconstruction techniques; (110.2350) Fiber optics imaging; (110.0113) Imaging through turbid media.*

Optical fibers are widely used in endoscopes for image transmission. Multimode fibers (MMFs) have attracted interest for minimally invasive medical imaging due to their ultra-thin shape and unparalleled information transport capabilities [1–4]. There has, however, been a great challenge to transport images through MMFs directly because of the strong mode coupling and interference within the fibers. When the light containing image information is coupled into a fiber, different modes with different phase velocities are excited and coupled with each other. As a result, a complex speckle pattern is formed at the fiber end face without any obvious features of the original image.

Due to mode dispersion, a MMF can be considered a disordered medium, for which the scattering effect can be compensated by modulating the phase and amplitude of the incident light field [5–7]. With the use of a liquid-crystal-based spatial light modulator (SLM) or digital micro-mirror device (DMD), the incident light field can be decomposed into sequential pixels. Then, by modulating the light field of each pixel, the desired light field can be transmitted through the fiber [5,8]. Particularly, Cizimar et al. used a wavefront optimization methods to achieve a raster-scan of a strongly focused light spot across the fiber end. By recording the intensity of the reflected or excited fluorescence light sequentially, both linear reflection and fluorescence imaging were achieved through a single MMF [1,9].

Digital phase conjugation (DPC) was also used for MMF-based imaging [10,11]. In a DPC system, a focused light spot is raster-scanned across the fiber end in advance, whilst the transmitted fields at the other end were holographically recorded with a reference light beam. By retrieving the phase from the captured holograms and projecting the phases onto SLM, the raster-scan of the focused light spot is achieved in its original positions. With the DPC method, Papadopoulos et al. used both continuous wave and pulsed laser sources, obtaining both fluorescence and photoacoustic images through single MMFs [10,11].

Popoff et al. pioneered calibrating the scattering effect of a disordered medium with a transmission matrix (TM) using phase-shifting interferometry [6]. Each TM element $t_{mn}$ is a complex coefficient linking the m[th] output pixel with the n[th] input pixel. With the measured TM, the phase mask on the SLM can be calculated to achieve focusing at the desired position at the distal fiber end. Raster-scanning the focused light spot was achieved for endoscopic fluorescence imaging [4]. Image transmission through an opaque medium was also achieved: the phase masks on the SLM were calculated using the captured transmitted light patterns and the prior measured TM [7]. With the measured TM, Loterie et al. developed a MMF-based confocal microscopy by digitally eliminating the noise from the distal end of fiber, which also achieved image reconstruction through a MMF [2]. Choi et al. calibrated a MMF by measuring the complex light fields of both incident and transmitted light beams. By retrieving the patterns reflected from the target and employing a speckle imaging method, images were reconstructed with high resolution [3]. Recently, reference-less methods were studied for diffuser and MMF calibration, allowing for simplification of the experimental system [12–14]. Neural networks were also implemented to overcome the highly variable nature of MMF channels to reconstruct an image using the output speckles [15,16].

In this letter, we demonstrated a single-shot image retrieval method through a MMF in a reference-less system using a genetic algorithm (GA). The image was shown at the proximal end of a fiber while a noise-like speckle pattern was captured at the distal end. The algorithm optimized a random mask with the same size as the image to produce a speckle pattern with high similarity to an experimentally captured pattern. We used the correlation coefficient of two patterns to evaluate the similarity. As a result, the image can be retrieved when the patterns have a similarity over 90%. Once the TM is calibrated, only one illumination is required for each image, which offers promise for single-shot imaging through a single MMF.

When a DMD fills the entrance pupil of a MMF, the transmitted field measured by a camera at the other fiber end is a linear superposition of the output fields from all DMD pixels. Here we define a DMD pixel as an 'input' pixel (total number N) and a camera pixel as an 'output' pixel (total number M). For the m$^{th}$ output pixel, the field $E_m$ is the superposition of all input fields $E_n$ multiplied by their corresponding TM elements $t_{mn}$:

$$E_m = \sum_{i=1}^{N} t_{mn} E_n \qquad (1)$$

For image retrieval, both the amplitude and phase of $E_m$ can be recovered from the hologram captured by the camera with the inclusion of a reference beam. The field of the input image can then be calculated as $E_n$. However, only intensity information of $E_m$ (the square of amplitude) is captured in a reference-less system, hence the input field $E_n$ cannot be calculated with Eq. 1 directly. Fortunately, the noise-like output speckles through the MMF contain the corresponding input image information. Here we used a GA method and the measured TM to retrieve images from their corresponding output speckles.

GAs are nature-inspired algorithms which evolve initial populations according to specific conditions, previously studied for focusing light through scattering media [17,18]. The principle of the GA used in this work is shown in Fig. 1. Firstly, a set of random DMD masks were generated in MATLAB as the initial generation. As the DMD provides binary amplitude modulation, we used '1' and '0' referring to turning the micro-mirrors 'ON' or 'OFF'. The intensity speckle patterns are calculated using the prior measured TM in Eq. 1. Note that the output is the intensity pattern rather than the light field in this diagram. Furthermore, while each input mask and the corresponding output pattern are shown as vectors in Fig. 1, experimentally these are shaped as square matrices. A cost function is used to estimate the quality of the output patterns. In this case, the experimentally obtained speckle pattern from the original image is used as the target, whilst the correlation coefficient of this target and the calculated output speckle pattern is used as the cost function, marked as CC1). The correlation between the initial mask and the optimized inoput mask is defined as CC2. These are realized with the corr2 function in MATLAB. Once the cost function of each intensity pattern is calculated, the patterns and corresponding input masks are ranked according to their correlation coefficient. Next, input masks are selected as parents to produce the next generation of masks. Input masks with a greater cost function have a higher chance of being selected as parents. The parent masks are then bred together, where random parts of one parent mask is replaced with the corresponding part of the other parent mask. The resulting mask then undergoes a random mutation. This is repeated, forming a new offspring generation with the same population as the initial generation. The offspring masks have their cost function measured and the process repeats. Therefore, the 'gene' responsible for input masks with high cost functions is saved and the population evolves to produce more of these masks with higher cost functions. After a specific number of iterations, an input mask is obtained with high similarity to the original image. Here we used the correlation coefficient of the original image and the optimized input mask to evaluate the quality of the retrieved image.

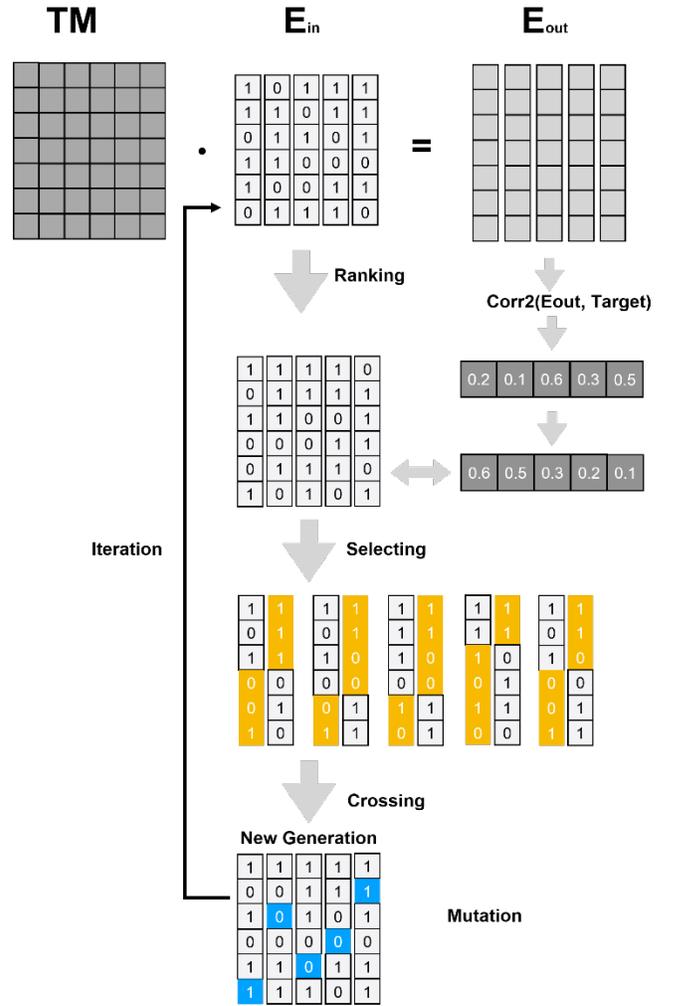

Fig. 1. Schematic diagram of the genetic algorithm for image retrieval. Each vector of '0' and '1' refers to a DMD mask. Input individuals are ranked according to their cost function with larger cost functions ranking higher, with a larger probability to be selected as parents for the next generation. In the crossing process, a random part of an input mask (yellow) is replaced with the corresponding part of the other mask in each parent pair. Subsequently, mutations are implemented at random positions (blue) to form the new generation for the iterative optimization process.

For clarity, this value is marked as CC2 while the correlation coefficient used as the cost function is marked as CC1 in the following.

The TM of a MMF was measured using the prVBEM algorithm in a DMD-based reference-less system [13]. The setup of the experiment is shown in Fig. 2. Laser light from a He-Ne laser (632.8 nm, 25-LHP-925-230, Melles Griot) was projected onto a DMD (1024×768 micro-mirrors, Discovery 1100, Texas Instruments) through a tube lens (AC254-200-A- ML, Thorlabs). The modulated fields were coupled into a straight MMF (50-μm-core, 30-cm-long, FG050UGA, Thorlabs) through a microscope objective (Nikon CFI Plan, Achro 20×, 0.65 NA, 0.56 mm WD, tube lens focal length: 200 mm). The transmitted light was magnified with a microscope objective (Olympus 20× Plan Achro, 0.4 NA, 1.2 mm WD, tube lens focal length: 180 mm) and a tube lens (AC254-200-A-ML, Thorlabs) before being captured by a CMOS camera (C1140-22CU, Hamamatsu). 2×2 micro-mirrors were combined as a single macro-pixel to enhance the difference between an 'ON' pixel and an 'OFF' pixel. We used N = 1296 macro-pixels for the input basis and M = 9216 pixels for the outputs. 8000 random DMD masks were generated by MATLAB at an 'ON'-'OFF' ratio of 50:50 and used as input masks. Different images were shown on the DMD whilst the transmitted light was captured by the same camera. In order to study the influence of bending on image retrieval, the middle point of the fiber was fixed onto a translation stage (RB13M/M, Thorlabs) and moved to different positions. Whilst the image projected onto the DMD remained the same, the intensity patterns captured by the camera changed with fiber bending. With the use of the straight fiber TM, image retrieval was implemented using different intensity patterns as targets.

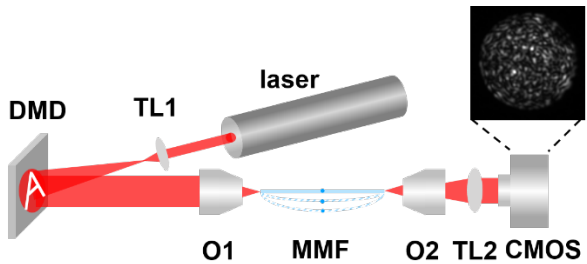

Fig. 2. Experimental configuration for TM calibration of a MMF and image retrieval. For TM calibration, 8000 random masks were shown on the DMD sequentially whilst the corresponding transmitted intensity speckles were captured by the CMOS camera. Once the TM was achieved, a figure was shown on the DMD and a noise-like speckle pattern was captured by the camera. The blue dashed lines refer to the bent MMF by translation of the middle point of the fiber. TL1, TL2 are tube lens 1 and 2; O1, O2 are objective 1 and 2; CMOS is the camera.

Image retrieval was implemented through the GA. After 60,000 iterations, images were retrieved from the noise-like intensity patterns. Some examples of retrieved images are shown in Fig. 3. The accuracy of image reconstruction improves proportional to iteration number (Fig. 4). After 60,000 iterations, CC1 was higher than 97% while the corresponding CC2 is over 73%. The features of the original image were recovered in the retrieved images. The computing time averaged 20 iterations per second in a PC (Intel Core i7); 60,000 iterations cost approximately 50 minutes.

Image retrieval was also implemented using the TM of the straight fiber and different intensity patterns captured through bent fibers. CC1 and CC2 decreased with increasing translation distance of the fiber middle point, as shown in Fig. 5. Features of the original image were distinguishable with small translational distances.

The experimental results demonstrate that arbitrary images can be retrieved from the intensity-only pattern transmitted through a MMF. Once the TM is measured, only a single illumination is needed to capture the intensity pattern through the fiber, which provides promise for single-shot imaging using a MMF in a simple reference-less system. The TM alone cannot be used for image retrieval, except for inputs used in its calculation. In contrast to raster-scanning imaging which is time-consuming due to the repeated modulation of an SLM [1,2], single-shot imaging improves the imaging rate significantly.

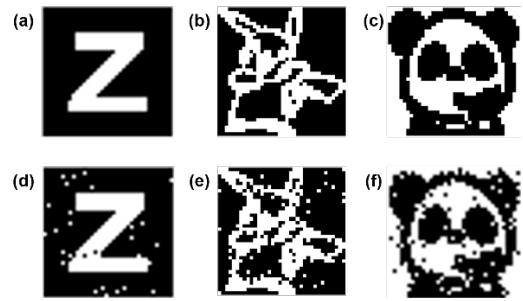

Fig. 3. Experimental results of image retrieval through a MMF. Different images of (a) the letter 'Z', (b) cartoon figure 'Pikachu' and (c) 'panda' are shown on the DMD. (d) (e) (f) are the corresponding retrieved images through the GA approach.

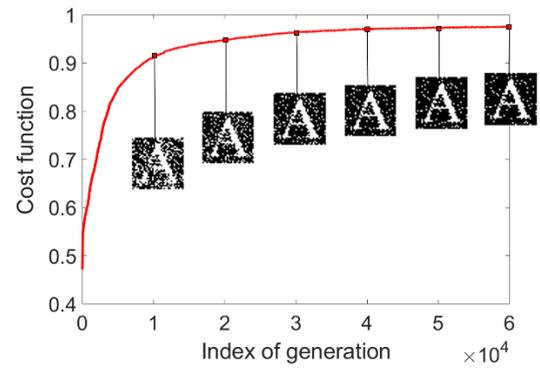

Fig. 4. The evolution of cost function (CC1) with iterations. The value of cost function of retrieved images in different generations from 10,000 to 60,000 are 91.25%, 94.79%, 96.36%, 97.04%, 97.33% and 97.62%, respectively. The correlation coefficient of original image and retrieved images are 39.06%, 52.72%, 61.43%, 67.24%, 70.16% and 73.21%, respectively.

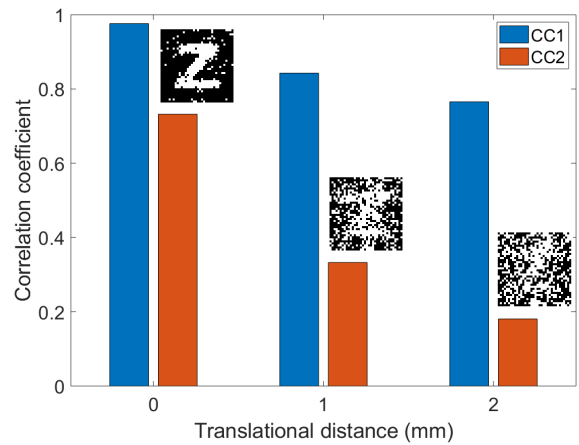

Fig. 5. The effect of translation of the middle of the fiber on CC1 and CC2 after 60,000 iterations. CC1 is the correlation coefficient of the calculated speckle pattern and the target pattern. CC2 is the correlation coefficient of the retrieved image and the original image.

The primary features of the original image can be identified after 10,000 iterations, while the features become clearer with the increase of CC2. The algorithm costs 10 minutes to retrieve an image from which the original image can be roughly identified. The optimization rate decreases with the increase in generation, and 50 minutes are required to retrieve an image with very clear features. In this case, an image contains 1296 pixels while an output pattern has 9216 pixels. The number of pixels is determined by the prVBEM algorithm used for reference-less TM calibration of the MMF. By using a holographic method for TM calibration, the number of pixels in the output patterns can be reduced, which in turn reduces the computing time for image retrieval. In addition, the computing time can further be reduced by using a GPU.

As mentioned above, the primary challenge for imaging with a MMF is the complex mode coupling and interference. Due to the complex phase delay and mode coupling, injecting an image at the input fiber end produces a noise-like pattern at the output end. Image transmission through a MMF was achieved by recovering the input light field from the output hologram using a measured TM [2,7].

In this work, only the intensity of both input and output light was used for image retrieval, suggesting that in a MMF-based light transport system, the main intensity features of the input light survive in the output intensity-only pattern in a specific modality. On the other hand, CC1 reached approximately 73% while the CC2 was approximately 97%. This suggests that there are some input pixels leading to similar output intensity patterns. This issue is also demonstrated in light focusing studies through scattering media [5,6,12], where modulating the input light field for focusing at a specific camera pixel also causes intensity enhancement surrounding the target pixel. This is because the TM for these output pixels has some similar features.

According to mode dispersion, the incident light is spatially distributed into different modes during transport through a MMF. The equivalent mathematical definition of a mode can be expressed by an electromagnetic field, and is the solution to Maxwell's wave equation, while in spatial wavefront shaping, each mode is considered as an independent pixel. In this case, each input mode refers to a DMD micromirror, which has a 'ON' or 'OFF' state. In TM theory, each input mode is connected to an output mode with a complex coefficient, which means that each input has a specific connection with an output, namely the amplitude and phase. This work suggests that each input also has a specific connection with an output not only in phase and amplitude, but also in intensity pattern, a combination of both. Every mode in the fiber has a specific phase change and carries image information with a given power density of light, and therefore each mode has a specific influence on the resulting output intensity pattern. As shown in Figure 5, the transmission of both light and phase changes with increasing fiber deformation, so both image information and power density have been exchanged between different modes because of coupling. This theory is also in accordance with our recent work on reference-less MMF calibration without phase retrieval [12].

In conclusion, we developed an image retrieval method through a MMF in a reference-less system using a genetic algorithm (GA). By using the correlation coefficient of an experimentally captured transmitted pattern and a calculated one, a random input mask can be optimized into an image with high similarity to the original binary DMD image. This offers promise for single-shot endoscopic imaging through an ultrathin MMF in a simple, reference-less system.

**Funding.** Engineering and Physical Sciences Research Council [grant number EP/L022559/1 and EP/L022559/2]; Royal Society [grant numbers RG130230 and IE161214].